\documentclass[noshowpacs,amsmath,
twocolumn,
superscriptaddress,
8pt,
aps,prb]{revtex4-1}
\usepackage{setspace}
\usepackage{amsmath}
\usepackage{graphicx}
\usepackage[nearskip,margin = 0pt]{subfig}

\usepackage{verbatim}
\usepackage{amsfonts}
\usepackage{amssymb}
\usepackage{epstopdf} 
\usepackage{xcolor}
\usepackage{verbatim}
\DeclareGraphicsExtensions{.pdf,.eps,.png,.jpg,.mps} \usepackage[utf8]{inputenc}
\begin{document}

\title{Towards high-power, high-coherence, integrated photonic mmWave platform with microcavity solitons}

\author{Beichen Wang$^{1,*}$, Jesse S. Morgan$^{1,*}$, Keye Sun$^1$, Mandana Jahanbozorgi$^1$, Zijiao Yang$^2$, Madison Woodson$^3$, Steven Estrella$^3$, Andreas Beling$^{1,\dagger}$ and Xu Yi$^{1,2,\ddagger}$ \\
\vspace{3pt}
$^1$Department of Electrical and Computer Engineering, University of Virginia, Charlottesville, Virginia 22904, USA.\\
$^2$Department of Physics, University of Virginia, Charlottesville, Virginia 22904, USA.\\
$^3$Freedom Photonics LLC,
Santa Barbara, California, USA.\\
$^{\ast}$These authors contributed equally to this work.\\
Corresponding authors: $^{\dagger}$andreas@virginia.edu, $^{\ddagger}$yi@virginia.edu.}




\begin{abstract}
\noindent {\bf Abstract:} Millimeter-wave (mmWave) technology continues to draw large interest due to its broad applications in wireless communications, radar, and spectroscopy. Compared to pure electronic solutions, photonic-based mmWave generation provides wide bandwidth, low power dissipation, and remoting through low-loss fiber. However, at high frequencies, two major challenges exist for the photonic system: the power roll-off of the photodiode, and the large signal linewidth derived directly from the lasers. Here, we demonstrate a new photonic mmWave platform by combining integrated microresonator solitons and high-speed photodiodes to address the challenges in both power and coherence. The solitons, being inherently mode-locked, are measured to provide 5.8 dB additional gain through constructive interference among mmWave beatnotes, and the absolute mmWave power approaches the theoretical limit of conventional heterodyne detection at 100 GHz. In our free-running system, the soliton is capable of reducing the mmWave linewidth by two orders of magnitude from that of the pump laser. Our work leverages microresonator solitons and high-speed modified uni-traveling carrier photodiodes to provide a viable path to chip-scale  high-power, low-noise, high-frequency sources for mmWave applications.
\end{abstract}

\date{\today}

\maketitle

\noindent {\bf Introduction}

\noindent Millimeter-waves (mmWaves) provide key advantages in communication bandwidth, radar resolution, and spectroscopy thanks to their high carrier frequencies \cite{cooper2008penetrating,kleine2011review,koenig2013wireless}. 
Photonic oscillators operate at frequencies of hundreds of THz, and the frequency of the electrical signal  produced by, e.g. heterodyne detection of two lasers, is only limited by the photodiode bandwidth. However, at mmWave frequencies, the output power of the photonic system suffers from the power roll-off associated with the photodiode's bandwidth. In terms of signal coherence, stabilizing the frequency difference of two lasers to a low frequency reference is challenging for mmWaves due to the high frequency.

\begin{figure*}[!ht]
\captionsetup{singlelinecheck=off, justification = RaggedRight}
\includegraphics[width=18.0cm]{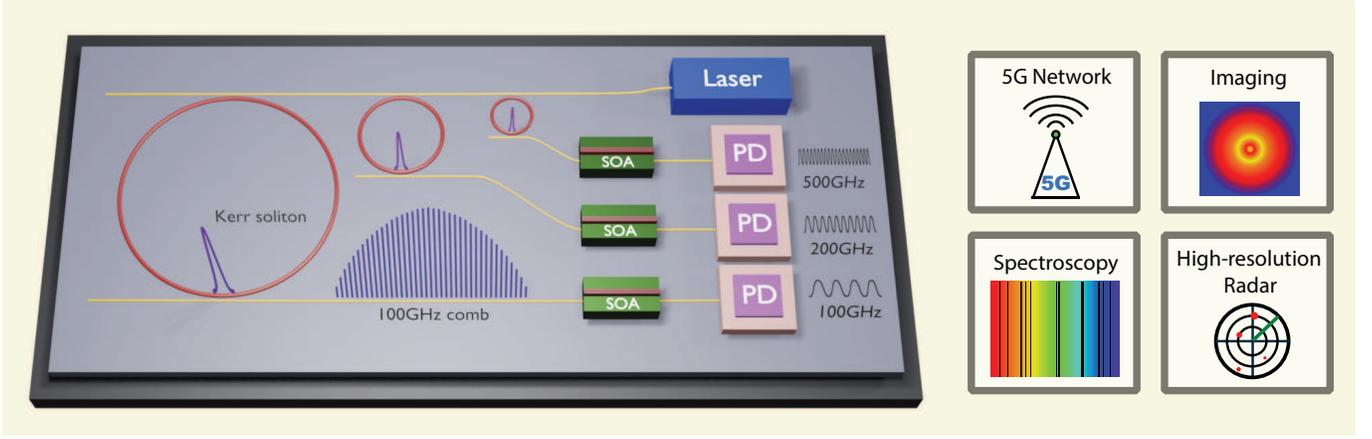}
\caption{{\bf Artistic conceptual view of fully integrated mmWave platform based on microresonator solitons.} The microresonator solitons are generated by pumping a high-Q Kerr microresonator with a continuous-wave (cw) laser. Photodetecting the solitons generates the mmWave signal at the soliton repetition frequency (comb spacing). Soliton mode-locking can provide up to 6 dB more power than that of conventional two laser heterodyne detection, and it is also capable of reducing the mmWave linewidth. By leveraging advances in photonic heterogeneous integration, all critical components, including pump laser, semiconductor optical amplifiers (SOAs) and ultrafast photodiodes (PDs), can potentially be integrated with the Kerr microresonators on the same chip. The integration will enable arrays of coherent mmWave sources, which can generate mmWave signals over a broad range of frequencies. Such a mmWave platform can advance applications in high-speed wireless communication, sub-THz imaging and spectroscopy, and high resolution ranging.}
\label{fig:count}
\end{figure*}

The recent development of dissipated Kerr solitons in microresonators\cite{herr2014temporal,yi2015soliton,brasch2016photonic,gong2018high,gaeta2019photonic,he2019self} provides an integrated solution to address the challenges of photonic-generated mmWaves in both power and coherence. These solitary wave packets achieve mode-locking by leveraging Kerr nonlinearity to compensate cavity loss and to balance chromatic dispersion \cite{herr2014temporal,kippenberg2018dissipative}. Microresonator solitons have been applied to metrology \cite{spencer2018optical}, optical communications \cite{marin2017microresonator} and spectroscopy\cite{suh2016microresonator,dutt2018chip} in the form of microresonator-based frequency combs (microcombs)\cite{del2007optical}. Due to the miniaturized dimension, the repetition rate of microresonator solitons ranges from a few GHz to THz \cite{suh2018gigahertz,li2017stably}. Direct detection of the solitons with a fast photodiode produces mmWave at the repetition frequency of the solitons. When compared with the conventional two laser heterodyne detection method, the soliton mode-locking provides up to 6 dB gain in mmWave output due to the constructive interference among beatnotes created by different pairs of neighboring comb lines \cite{kuo2010spectral}. This additional gain is of great importance at high frequencies, since it can relax the bandwidth  requirements  in the photodiode. In terms of signal coherence, recent studies have shown that the phase noise of the soliton repetition frequency at 10's of GHz can be orders of magnitude smaller than that of its pump laser \cite{liang2015high,yi2015soliton,yi2017single,liu2020photonic}. When microresonator solitons are married with integrated lasers \cite{stern2018battery,xiang2020narrow}, amplifiers \cite{de2020heterogeneous}, and high-speed photodiodes \cite{yu2020heterogeneous} through heterogeneous or hybrid integration, a fully integrated mmWave platform can be created with high power, high coherence performance and the potential for large scale deployment through mass production (Fig. \ref{fig:count}).

In this letter, we demonstrate high power, high coherence photonic mmWave generation at 100 GHz frequency through the combination of integrated microresonator solitons and a modified uni-traveling carrier photodiode (MUTC PD). A 5.8 dB increase of mmWave power is obtained by using microresonator solitons when comparing to the output power of conventional heterodyne detection. Importantly, the power level we achieve with microresonator solitons is approaching the theoretical limit of heterodyne detection, which assumes an ideal photodiode with zero power roll-off in its frequency response. The system also achieves a maximum mmWave power of 7 dBm, one of the highest powers ever reported at 100 GHz \cite{100Gbenchmark}. For our free-running system, the 100 GHz signal has Lorentzian and Gaussian linewidth of 0.2 kHz and 4.0 kHz, respectively, which is two orders of magnitude smaller than that of the pump laser. The dependence of output power on the number of comb lines and chromatic dispersion is carefully studied both theoretically and experimentally. Our demonstration paves the way for a fully integrated photonic microwave system with soliton microcombs and high-speed photodiodes. 

\noindent {\bf Results}

In conventional heterodyne detection, mmWaves are generated when two laser lines mix with each other on a photodiode and create one beat note. However, when using an optical frequency comb, each comb line will beat with its two adjacent neighbour lines to create beatnotes at the comb repetition frequency. For a comb that consists of $N$ comb lines, $(N-1)$ beat notes will be created at the comb repetition frequency. Therefore, for the same average optical power, the comb can produce up to twice the number of beatnotes per laser line than heterodyne detection, and thus generate twice the AC photocurrent. The output power from the photodiode at the comb repetition frequency can be described as \cite{FundPhoton,kuo2010spectral}:
\begin{align}
P_{PD} = \frac{I_{DC}^{2}R_{L}}{2} \left[\frac{2(N - 1)}{N}\right]^2 \times \Gamma,
\label{eqn:combCalc}
\end{align}

\noindent where $I_{DC}$ is the average photocurrent, $R_{L}$ (50 $\Omega$) is the load resistor, and $N \geq 2$ is the number of comb lines. $\Gamma$ is the measured relative mmWave power roll-off for the photodiode, and is $\sim5.5$ dB for the 7 $\mu m$ and $\sim6$ dB for the 8 $\mu m$ diameter PDs used in this work at 100 GHz. Clearly, the power at the limit of $N \to \infty$ is 4 times (6 dB) higher than the power of heterodyne detection, where $N=2$.

In practice, however, conventional frequency combs are not the best candidates to achieve the 6 dB gain for mmWave generation due to their low repetition frequencies. Previously, two attempts with electro-optics modulation frequency combs were reported, where line-by-line amplitude and phase shaping was used to remove the unnecessary comb lines and increase the repetition rate from 20 GHz to 100 and 160 GHz \cite{kuo2010spectral,wun2014photonic}. This post spectral filtering nonetheless increases the complexity and cost of the system. Conversely, microresonator solitons have comb repetition rates ranging from a few GHz to 1 THz, and can be directly applied to mmWave generation. MmWave generation with soliton microcombs in tapered-coupled microtoroid resonator \cite{zhang2019terahertz}, from dual-comb structure \cite{zang:2020}, and from a pair of comb lines \cite{tetsumoto2020300} have been shown, but there was no investigation into the output power.

\begin{figure*}[!ht]
\captionsetup{singlelinecheck=off, justification = RaggedRight}
\includegraphics[width=18.0cm]{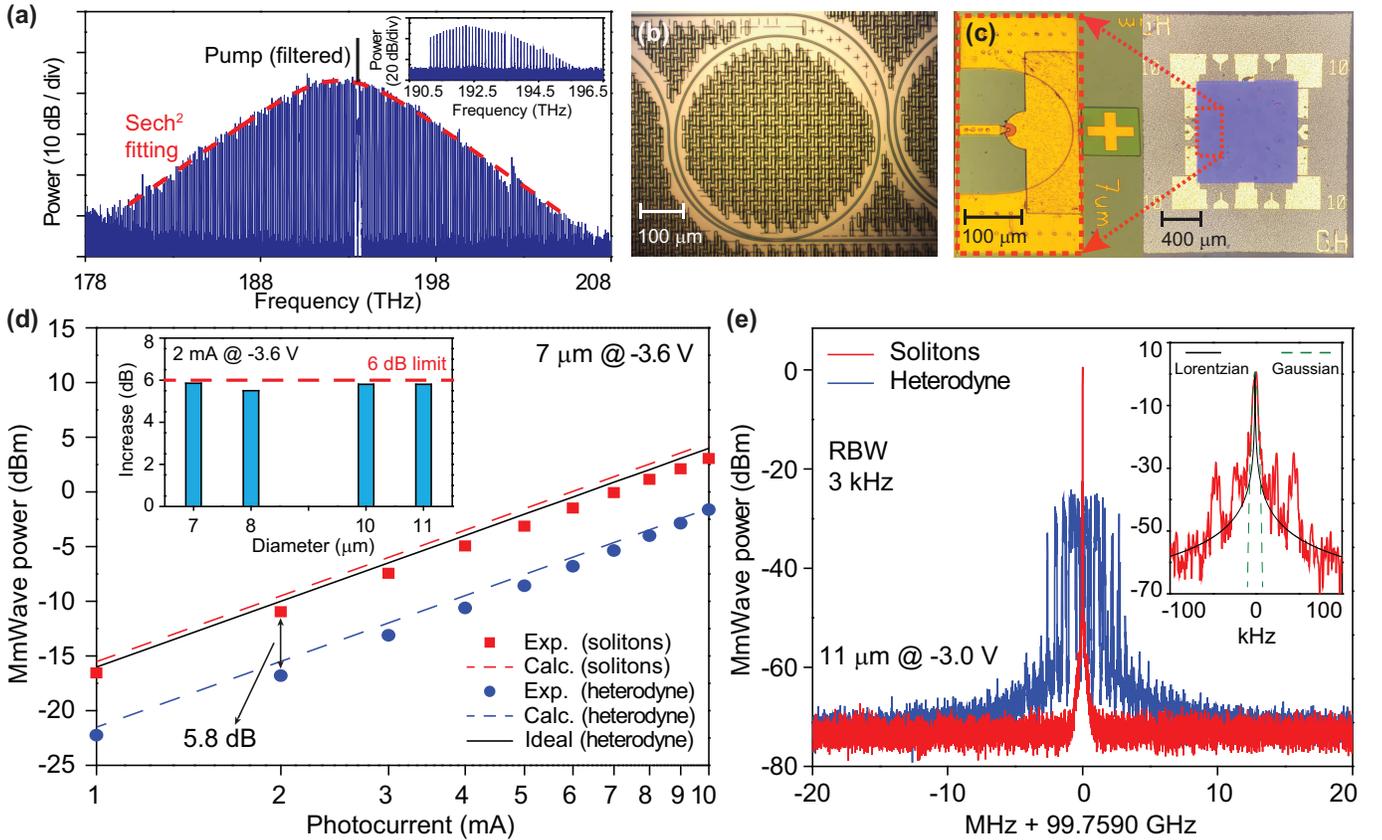}
\caption{{\bf Summary of featured experimental data of 100 GHz mmWave generation.} {\bf(a)} Optical spectrum of single soliton state from the microresonator. The spectrum has sech$^2$ spectral envelope (fitting shown in dashed red line). The pump laser is suppressed by a fiber Bragg grating filter. Inset shows the optical spectrum of soliton frequency comb after amplification and dispersion compensation. {\bf(b)} Microscopic image of integrated Si$_3$N$_4$ microresonator with 100 GHz free spectral range (FSR). {\bf(c)} Microscopic images: front of photodiode die zoomed in on single 7$\mu$m device (left), and back of photodiode die flip-chip bonded to aluminum nitride submount (right). {\bf(d)} 100 GHz mmWave output power measured for microresonator solitons (red) and optical heterodyne detection of two cw-lasers (blue). The mmWave output power from the soliton is $\sim 5.8$ dB more than that of the heterodyne detection at the same photocurrent. Theoretical calculated powers from equation (1) are shown in dashed lines. Particularly, ideal output power from heterodyne detection is illustrated with black solid line, which serves as a theoretical limit of heterodyne detection assuming no PD power roll-off at 100 GHz frequency. The inset shows the power increase by using solitons over optical heterodyne on four devices with different diameters. {\bf(e)} Down-converted electrical spectrum of 100 GHz signal generated with free-running microresonator solitons (red). Inset shows the fitting with Lorentzian (black) and Gaussian (dashed green) lineshapes and the corresponding 3-dB linewidths are 0.2 kHz and 4 kHz respectively. As a comparison, the signal generated from heterodyne method is shown in blue trace. The PD diameter and bias voltage are indicated in each panel.}
\label{fig:data1}
\end{figure*}

\begin{figure*}[ht!]
\captionsetup{singlelinecheck=off, justification = RaggedRight}
\includegraphics[width=18.0cm]{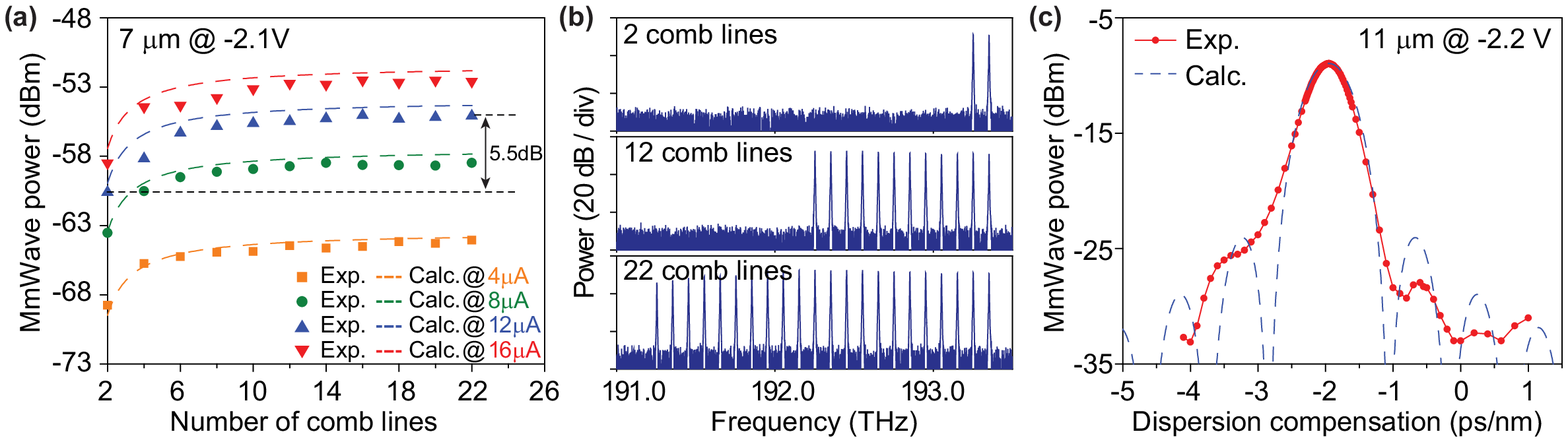}
\caption{{\bf MmWave power versus number of comb lines and dispersion.} {\bf(a)} MmWave power at 100 GHz for different number of comb lines at four different photocurrents. The measurements agree very well with the theoretical calculation based on equation (1), which are shown in dashed lines. {\bf(b)} Corresponding optical spectra of two, twelve and twenty-two comb line measurements in panel (a). {\bf(c)} MmWave power versus dispersion compensation added by waveshaper, $d_c$. The maximum output power is reached at $d_c =-1.95$ ps/nm, where the dispersion from fiber and EDFA is completely compensated. A theoretical curve from equation (2) is shown in dashed line and agrees very well with the measurement. The PD diameter and bias voltage are indicated in each panel.}
\label{fig:data2}
\end{figure*}

The dissipated Kerr solitons used in this work are generated in an integrated, bus-waveguide coupled Si$_{3}$N$_{4}$ micro-ring resonator with free spectral range (FSR) of $\sim$100 GHz. The single soliton state with a 35.4 fs pulse width is generated and its squared hyperbolic secant spectral envelope is characterized by an optical spectrum analyzer (Fig. 2\textbf{a}). The comb is then amplified by an erbium-doped fiber amplifier (EDFA) and sent to the photodiode, and an optical programmable waveshaper (WS) is used to compensate the group velocity dispersion and to suppress spontaneous emission (ASE) noise from the EDFA. The inset of Fig. 2\textbf{a} shows the optical spectrum after the amplification and dispersion compensation. The photodiode used in this work is based on the charge-compensated modified uni-traveling carrier photodiode (MUTC PD) structure. MUTC PDs operate under the principle of single carrier transit, and compared to traditional p-i-n photodiodes, isolating electrons for this transit process eliminates the dependency on the slower-traveling holes leading to higher-speed operation. To further enhance performance and limit thermal degradation, the PDs are then flip-chip bonded to a ceramic substrate made of gold transmission lines grown on aluminum nitride submount (AlN) \cite{AlN}. Pictures of the microresonator and a PD die are shown in Fig. 2\textbf{b} and Fig. 2\textbf{c}, respectively. Details of microresonator solitons and photodiodes are described in the Materials and Methods section.

To characterize the 6 dB power increase from the microresonator solitons, the PD output powers are measured for both microresonator soliton detection and heterodyne detection on four of our PDs with 7, 8, 10, and 11 $\mu m$ diameters. The heterodyne measurements are performed using two continuous-wave lasers with the same optical power and polarization. A variable optical attenuator is used to control the optical power illuminating on the PD. In the linear region of PD operation, the 100 GHz mmWave powers at different photocurrents are shown in Fig. 2\textbf{d} for the 7-$\mu m$ device. The DC photocurrent is a direct measurement of the optical power illuminating on the PD. In the experiment, the coupling distance from fiber to PD is increased for a uniform illumination, resulting in 1 mA photocurrent for 11 mW optical input power. The mmWave power generated from the microresonator solitons is measured to be 5.8 dB higher than that of heterodyne detection. This power increase is approaching the 6 dB theoretical limit, and is verified on all four PDs with different diameters (shown in the inset of Fig. 2\textbf{d}). As a result of the 6 dB power increase, the mmWave power generated using microresonator solitons is within 1 dB of the theoretical power limit of heterodyne detection (solid black line in Fig. 2\textbf{d}), where the detector is assumed to be ideal and has no power roll-off at mmWave frequency. It shall be noted that no optical spectrum flattening is applied in our measurement. For 5.8 dB power improvement, a 3 dB bandwidth of 7 comb lines is required for the Sech$^2$ or Gaussian spectral envelope. As discussed in the Materials and Methods section, the shape of the spectral envelope has little effect on mmWave power when the number of comb lines is large.

The electrical spectrum of the 100 GHz mmWave signal is measured and shown in Fig. 2\textbf{e}. Limited by the available bandwidth of our electrical spectrum analyzer, we down convert the 100 GHz mmWave by sending it to an RF mixer to mix it with the fifth harmonic of a 20.2 GHz local oscillator. The mixer generates a difference frequency at $\Delta f = 5f_{LO} - f_r$. $\Delta f$ is measured to be 1.2410 GHz, and we can derive the mmWave frequency as $f_r = 99.7590$ GHz. A low-noise, narrow signal is clearly observed at 3 kHz resolution bandwidth (RBW) in Fig. \ref{fig:data1}\textbf{e} (red trace). The signal is fitted with a Lorentzian, and the 3-dB bandwidth is $0.2$ kHz (zoomed-in panel in Fig. \ref{fig:data1}\textbf{e}). Note that the soliton repetition rate is subject to fluctuations (laser frequency drift, temperature, etc.), and the central part of the signal is Gaussian with 3-dB linewidth of $4$ kHz. This narrow linewidth at 100 GHz frequency is obtained for a free-running microcavity soliton, which is driven by a pump laser with significantly broader linewidth ($\sim 200$ kHz, New Focus 6700 series specification). To compare the signal coherence between conventional heterodyne method and the soliton method, the heterodyne signal of beating the pump laser and another 6700 series New Focus laser is also measured and shown in Fig. \ref{fig:data1}\textbf{e} (blue trace). At the same RBW, the heterodyne signal has poor coherence and its frequency is drifting $> 5$ MHz. Our measurements show that using free-running microcavity solitons can reduce the linewidth of mmWave signals by 2 orders of magnitude, giving the microresonator soliton platform a key advantage over conventional heterodyne detection. No RF reference is used to stabilize the mmWave; in fact, the only controls used are the coarse temperature controls of the laser and the microresonator, used to offset the change in environmental temperature. Further measurements of phase noise and Allan deviation will be introduced in later paragraphs.

\begin{figure*}[!t]
\captionsetup{singlelinecheck=off, justification = RaggedRight}
\includegraphics[width=18cm]{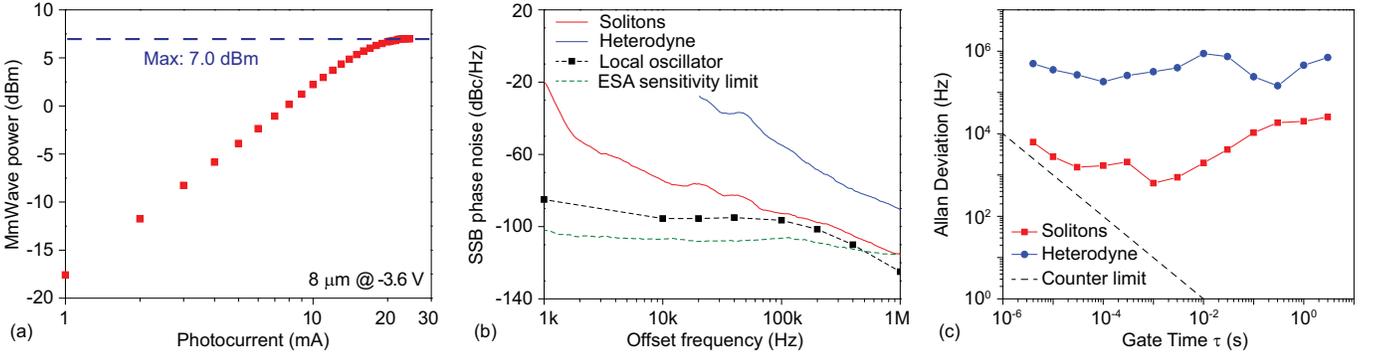}
\caption{{\bf Measurement of mmWave power, mmWave phase noise and Allan deviation.} {\bf (a)} Maximum power of 7 dBm is reached at 22.5 mA and $-3.6$ V bias voltage in the $8\mu m$ device. {\bf (b)} Phase noises of the free-running soliton-based mmWave (red) and the heterodyne mmWave (blue) at 100 GHz. The measurement sensitivity floor is set by both the ESA sensitivity limit (dash green), and the local oscillator phase noise (dash black). {\bf (c)} Allan deviation of the free-running soliton-based mmWave (red) and the heterodyne mmWave (blue).}
\label{fig:data3}
\end{figure*}

Next we verify the dependence of mmWave power increase on the number of comb lines, which is described in equation (1). A line-by line waveshaping filter is used to select the number of comb lines that pass to the MUTC PD. We test the number of comb lines from 2 to 22 at four different photocurrent levels (optical power), and the result is shown in Fig. 3{\bf a}. Three representative optical spectra for 2, 12, and 22 comb lines are shown in Fig. 3{\bf b}. The measured mmWave power follows the calculated curves. Interestingly, a 3 or 5 dB increase of power only requires 4 or 9 comb lines. This relatively small demand for comb lines relaxes the microresonator soliton requirement in terms of its optical bandwidth. 

The increase of mmWave power only happens when the beatnotes generated by different pairs of comb lines are in constructive interference. This is not always the case if there is dispersion between the microresonator and the PD. This effect is studied by applying programmable dispersion using a waveshaper. The measurement of mmWave power versus waveshaper dispersion is shown in Fig. 3{\bf c}. The effect can be calculated analytically by adding phase to each comb line, and will modify equation (1) to: 

\begin{equation}
P_{PD} = \frac{I_{DC}^{2}R_{L}}{2}\left[\frac{2\sin\left[(N-1)\pi c d f_r^2/f_p^2\right]}{N\sin\left[\pi c d f_r^2/f_p^2\right]}\right]^{2}\times \Gamma,
    \label{eqn:dispersion}
\end{equation}{}

\noindent where $c$ is the speed of light, and $d =d_0 + d_c$ is the accumulated group velocity dispersion between the microresonator and PD. $d_0$ denotes the offset dispersion in the system introduced by fibers and amplifiers, and $d_c$ represents the dispersion compensation added by the waveshaper. The derivation of equation (2) is shown in the Materials and Methods section. The measurement and theory prediction agree very well when an offset dispersion of $d_0 =1.95$ ps/nm is included. The offset dispersion exists in our system because of the 70 meter fiber used to connect the microcomb lab and photodetector lab (contributing 1.26 ps/nm), with the rest of the dispersion coming from the fibers in the EDFA. $N$ is used as a free parameter for fitting the experimental curve, and $N=15$ is used for the dashed line in Fig. 3{\bf c}. The fitted $N$ should be interpreted as the effective number of comb lines to account for the spectral envelope shape. When the entire system is fully integrated, the overall length of waveguides will be well below a meter, and the dispersion will not impact the mmWave power.

\begin{figure*}[!hbt]
\captionsetup{singlelinecheck=off, justification = RaggedRight}
\includegraphics[width=18.0cm]{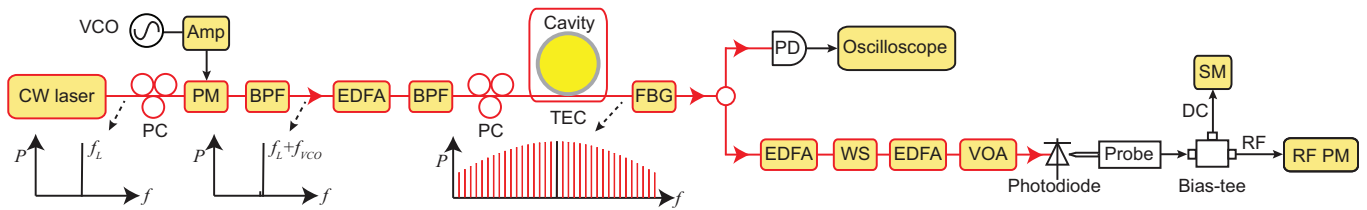}
\caption{{\bf Experimental setup.} The microresonator solitons are generated in a SiN resonator which is coarsely temperature controlled by thermoelectric cooler (TEC). The pump laser is the first modulation sideband of a phase modulated (PM) continuous wave (cw) laser, and the sideband frequency can be rapidly tuned by a voltage controlled oscillator (VCO). The frequencies of the cw laser and phase modulation are $f_L$ and $f_\text{VCO1}$, respectively. The pump laser is then amplified by an erbium-doped fiber amplifier (EDFA), and the amplified spontaneous emission noise is filtered out by a bandpass filter (BPF). At the output of the resonator, a fiber-Bragg grating filter is used to suppress the pump. The microresonator solitons are then amplified, dispersion compensated by a waveshaper (WS), and sent to the photodiode. The configuration also includes polarization controllers (PC), variable optical attenuator (VOA), source meter (SM), and RF power meter (RF PM).}
\label{fig:setup}
\end{figure*}

We obtain a maximum output power of 7 dBm at 22.5 mA for the 8 $\mu m$ device shown in Fig. \ref{fig:data3}a, due to the optimized light coupling from the size match of the 8 $\mu$m spot-size collimated fiber and diameter of the PD's absorber.
Using equation (1) we find that the ideal heterodyne response for this 8 $\mu m$ device would need 26.7 mA to achieve 7 dBm, which means we can produce the same power at lower average photocurrent using soliton excitation. The 7 dBm saturation power is recorded at -3.6 V bias. Increasing the reverse bias  can improve the saturation power, however, ultimately this can cause PD thermal failure \cite{xie2015photonic}, which is due to the raise in junction temperature from the dissipated power in the PD (reverse bias $\times$ average photocurrent).  One advantage of using solitons is that they can generate the same RF output power at a lower photocurrent than the two-laser heterodyne method, and thus can reduce the dissipated power and allow the PD to be operated further below  the point of thermal failure.

We further characterize the phase noise of the mmWaves generated from the free-running microcavity solitons, and compare it to the phase noise from the heterodyne method. Similar to the linewidth measurement, the 100 GHz mmWave signal is down converted in an RF mixer where it is mixed with the fifth harmonic of a 20.2 GHz local oscillator. To minimize the effect of frequency drifting in the phase noise measurement, the frequency of the down-converted signal is further divided down electrically by a factor of 14 and 100 for the soliton and heterodyne, respectively. The phase noise is then measured in the electrical spectrum analyzer with direct detection technique, and the result (at 100 GHz) is shown in Fig. \ref{fig:data3}b. Due to the large frequency drift, the heterodyne phase noise below 20 kHz offset frequency cannot be accurately characterized and thus is not presented. The soliton phase noise beyond 100 kHz is potentially limited by the measurement sensitivity, which is set by the noise floor of the spectrum analyzer (dash green), and the phase noise of the local oscillator (Keysight, PSG E8257D) used to down-convert the mmWave (dash black). The measurement shows that the free-running solitons can reduce the mmWave phase noise by $>$ 25 dB from the heterodyne method. The reduction of phase noise from the pump laser frequency to the soliton repetition rate is a result of the noise transfer mechanism in microresonator solitons \cite{yi2017single}. Our observation is in agreement with the previous reports for microresonator solitons at X-band and K-band repetition frequencies \cite{yi2017single,liu2020photonic}. The phase noise of soliton-based mmWaves can be further reduced in the future by using a pump laser with higher stability\cite{lucas2020ultralow}, tuning the soliton into quiet operation point \cite{yi2017single}, and implementing better temperature control of the entire system. For instance, compact external-cavity diode laser has achieved Lorentzian linewidth of 62 Hz recently\cite{volet2018micro}. Using this laser to drive the soliton could further reduce the free-running mmWave phase noise.

Finally, the Allan deviations of the mmWave generated from the soliton and the heterodyne detection are measured by counting the frequency of the down-converted signal on a zero dead-time counter (Fig. \ref{fig:data3}c). At 1 ms gate time, the Allan deviation of the soliton-based mmWave reaches the minimum of $<$ 0.7 kHz, which is more than two orders of magnitude better than that of the heterodyne detection. 
Above 1 ms gate time, the Allan deviation of the soliton-based mmWave increases due to the drift of pump laser frequency and temperature fluctuation. Stabilizing the mmWave signal to a low frequency reference could provide long term stability, which will increase the system complexity, but is possible through the electro-optics modulation method \cite{tetsumoto2020300}, or dual microcavity soliton methods \cite{spencer2018optical,wang2020vernier}.




\noindent {\bf Discussion}

In summary, we have demonstrated high-power, high-coherence mmWave generation at 100 GHz by using integrated microresonator solitons and MUTC PDs. Extending the frequency to several hundred GHz is possible. For the microresonator solitons, the highest repetition rate reported is 1 THz \cite{li2017stably}, while demonstrated MUTC PDs have detection capabilities of at least 300 GHz \cite{145G,dulme2019300}. As the microresonator solitons consume very little pump power, and most of the pump transmits through the waveguide \cite{yi2015soliton}, it is possible to recycle the pump laser power to drive the next microresonator solitons (Fig. 1). Two tandem microresonator solitons driven by the same pump laser have been reported previously \cite{dutt2018chip}. The proposed platform has the potential to be fully integrated on a single chip which can enable large-scale  mmWave arrays. The four critical components: laser, Kerr microresonator, amplifier, and ultrafast photodiode, have all been shown to be compatible with Si$_3$N$_4$ photonic platforms through heterogeneous integration. Once all components are fully  integrated, we expect that the platform can deliver a new paradigm regarding scalable, integrated photonics technologies for applications at very high frequencies, and thus provide  a path to compact, low-noise high-frequency sources for spectroscopy, ranging, and wireless communications.
\\

\noindent\textbf{Materials and Methods}

\begin{footnotesize}
\noindent{\bf Microresonator soliton generation.}
The dissipated Kerr solitons used in this work are generated in an integrated, bus-waveguide coupled Si$_{3}$N$_{4}$ micro-ring resonator. The resonator has a free spectral range (FSR) of $\sim$100 GHz, and an instrinsic quality factor of $2.6\times10^6$ and loaded quality factor of $2.2\times10^6$. The SiN resonator has a cross-section, width$\times$height, of $1.65 \times 0.8$ $\mu$m$^2$, and is coupled to a bus-waveguide of the same cross-section. The resonator radius is 0.24 $\mu$m, and the soliton-generation mode has anomalous dispersion of $\sim 1$ MHz/FSR. To generate a single soliton state, a rapid pump laser frequency scanning method \cite{stone2018thermal} is applied to overcome the thermal complexity when accessing the red-detuned soliton existence regime. The detailed experimental setup is shown in Fig. 5. The pump laser is derived from the first phase modulated sideband of a continuous wave laser, and the sideband frequency can be rapidly tuned by a voltage controlled oscillator (VCO). The pump laser scans its frequency at the speed of $\sim$20 GHz/$\mu$s, and the scan is stopped immediately once the pump laser frequency reaches the red-detuned regime of the resonator. The optical spectrum has a 3-dB bandwidth of $5.4$ THz, which contains a sufficient number of comb lines for photodetection. A thermoelectric cooler (TEC) is placed beneath the microresonator to coarsely overcome the environmental temperature fluctuations.
\medskip

\noindent{\bf Modified uni-traveling carrier photodiode.}
The charge-compensated modified uni-traveling carrier photodiode (MUTC PD) operates under the principle of single carrier transit, and compared to traditional p-i-n photodiodes, isolating electrons for this transit process eliminates the dependency on the slower-traveling holes leading to higher-speed operation. To accomplish this, the photon absorption process which generates electron-hole pairs in the PD absorber layer, occurs close to the p-contact layer allowing the excess holes to be quickly collected in response to the p-type material dielectric relaxation time. To further enhance the speed of the PD response, by step grading the doping of the partially-depleted absorber, an electric field is generated which accelerates the electrons through the absorber and towards the transparent and depleted drift layer. To prevent electric field collapse at the heterointerface of the absorber and drift layer, a fully-depleted absorber layer and a moderately doped cliff layer help to maintain electric field strength and accelerate the electrons into the drift layer \cite{cliff01,cliff02}. Once in the drift layer, electron space-charge effects are mitigated or charge-compensated by the light n-type doping in the drift layer \cite{n-dope}. 
Fabrication flow of the PDs and similar PD epitaxial layering structures have been reported previously \cite{ccmutc}, and so has the AlN submount \cite{160G}. The MUTC PDs used in this experiment have demonstrated dark currents as low as 200 pA at -2 V, 3-dB bandwidth of up to 145 GHz (4-$\mu m$ diameter PD), responsivity of 0.2 A/W at 1550-nm, and -2.6 dBm maximum output power at 160 GHz at -3 V bias \cite{145G}. 
They have also been investigated as viable receivers for soliton applications ranging from 50-500 GHz \cite{solitonProposed01,solitonProposed02}. The 3-dB bandwidth at 5 mA and -3 V bias for the 7, 8, 10, and 11-$\mu m$ diameter PDs used in this experiment are 92 GHz, 90 GHz, 70 GHz, and 70 GHz, respectively. Note that in Fig. 2{\bf d} ideal heterodyne power calculated using equation (1) where $N=2$, assumes 100\% modulation depth; however, measured modulation depth of the signal was 89\% leading to the observed mismatch in measured and calculated heterodyne power.

\medskip
\noindent{\bf MmWave linewidth reduction.}
Our observation of linewidth reduction is in agreement with previous reports of microresonator solitons at X- and K-band repetition frequencies \cite{yi2017single,liu2020photonic}. The soliton repetition frequency equals to the cavity free-spectral range (FSR) at the wavelength of soliton spectral envelope center. Both Raman self-frequency shift \cite{karpov2016raman} and dispersive wave recoils can affect the soliton envelope center wavelength \cite{brasch2016photonic,yi2017single}, and they are functions of laser-cavity frequency detuning. This can be clearly seen in Fig. 2\textbf{a}, as our soliton's envelope center is to the red side of the pump laser. Because of the chromatic dispersion, the FSR at different wavelengths is different, and thus the variation of the pump laser frequency, $f_p$, will alter the soliton spectral envelope center, and change the soliton repetition rate, $f_r$. To the first order, the transfer of frequency variation from the pump ($\delta f_p$) to the repetition rate ($\delta f_r$) can be described as $\delta f_r = \frac{\partial f_r}{\partial f_p} \times \delta f_p$, where $\delta$ denotes the variation. For both silica and silicon nitride resonators, this transfer coefficient $\frac{\partial f_r}{\partial f_p}$ has been measured to be on the level of $10^{-2}$ \cite{yi2017single,bao2017soliton}, and thus the soliton repetition rate linewidth is much smaller than that of the pump laser. The characterization of phase noise reduction from pump laser frequency to repetition frequency is currently unavailable in our system, as the frequency drift of our pump laser is too large for phase noise measurement. 

\medskip
\noindent{\bf MmWave power versus dispersion.} Optical pulses that propagate in an optical fiber will acquire addition phase due to group velocity dispersion in the fiber. Suppose the center frequency of the pulse is $\omega_p$, then the component at frequency $\omega$ will acquire a relative phase after propagation of distance $z$ \cite{agrawal2007nonlinear}:
\begin{equation}
{E}(z,\omega) = E(0,\omega)\exp [-i\frac{D_{\lambda}\lambda^{2}}{4\pi c}(\omega-\omega_p)^{2}z ] + c.c.,
\end{equation}
where ${E}(0,\omega) =E_0/\sqrt{2N} \exp(-i \omega t) $ is the electrical field of light at frequency $\omega$ and position $z = 0$, normalized to the photon number per unit time. Here we have assumed a flat spectrum for the comb, and $N$ as the total number of comb lines. $D_{\lambda}$ is the group velocity dispersion parameter, and $D_{\lambda} \approx 18$ ps/nm/km for SMF-28 fiber at 1550 nm. For soliton frequency combs, $(\omega-\omega_p)/2\pi = n\times f_r$ for the $n$-th comb line from the spectral envelope center, where $f_r$ is the comb repetition frequency.  Therefore, the photocurrent generated in the photodiode is
\begin{equation}
\begin{aligned}
I \equiv I_{DC} + I_{AC} =  |E|^2 = |\sum_{-N_0}^{N_0}E(0,\omega)\exp [\frac{-i \pi c f_r^2}{f_p^2} n^2 D_{\lambda}z ]+ c.c.|^2 
\\
=|E_0|^2 + |E_0|^2 \frac{2\sin\left[(N-1)\pi c D_{\lambda}z f_r^2/f_p^2\right]}{N\sin\left[\pi c D_{\lambda}z f_r^2/f_p^2\right]}\cos{(2\pi f_{r}t)} + ...,
\end{aligned}{}
\label{eqn:photocurrent}
\end{equation}
where we have used the $\sum_{k=m}^{n}ar^{k}=a(r^{m}-r^{n+1})/(1-r)$ to derive the term of $\cos{(2\pi f_{r}t)}$, and we have set $2 N_0+1 = N$. Higher harmonics of the repetition frequency are neglected as they are beyond the detection limit of our photodiode. 
Considering $I_{DC}$ as the average photocurrent flowing through the load resistor $R_{L}$, the detected mmWave power at frequency $f_r$ is yielded as:

\begin{equation}
P_{f_r} = \frac{I_{DC}^{2}R_{L} \Gamma}{2}\left[\frac{2\sin\left[(N-1)\pi c d f_r^2/f_p^2\right]}{N\sin\left[\pi c d f_r^2/f_p^2\right]}\right]^{2},
\label{eqn:RFpower}
\end{equation}
where we have defined $d=D_{\lambda}z$ as accumulated dispersion, and $\Gamma$ is the PD power roll-off at the repetition frequency. This equation is the same as equation (2) in the main text.
\noindent When dispersion is very small $(d\to 0)$, the detected mmWave power approximates to

\begin{equation}
P_{f_r} = \frac{I_{DC}^{2}R_{L}\Gamma}{2}\left[\frac{2(N-1)}{N}\right]^{2},
\label{eqn:RFpower2}
\end{equation}
which is equation (1) in the main text.

\medskip
\noindent{\bf MmWave power versus optical spectral envelope.} In this section, we calculate the impact of the optical spectral envelope on the mmWave power. For simplicity, we assume that the optical envelope is symmetric along the envelope center, and we assume no accumulated dispersion. For the $n$-th comb line, we have:
\begin{equation}
{E}(\omega_n) =f(n) \frac{E_0}{\sqrt{2N}} e^{-i \omega_n t} + c.c.,
\end{equation}
where function $f(n)$ is real and it describes the spectral envelope. We focus on the case where the number of comb lines is large, so that we can assume the envelope is smooth, and $|f(n+1)-f(n)| \ll f(n)$. The photocurrent is then expressed as:

\begin{equation}
\begin{split}
I =  |E|^2 = |\sum_{-N_0}^{N_0}f(n) \frac{E_0}{\sqrt{2N}} e^{-i \omega_n t} + c.c.|^2 
=\frac{|E_0|^2}{N} \sum_{-N_0}^{N_0}f^2(n)
\\+ \frac{2|E_0|^2}{N}\cos{(2\pi f_{r}t)} \times \sum_{n=-N_0}^{N_0-1}f(n)f(n+1) + ...,
\end{split}{}
\label{}
\end{equation}
where we have neglected higher harmonics of the repetition frequency again. The sum can be simplified by using the symmetric envelope condition, $f(-n) = f(n)$, and we can substitute $f(n+1) = f(n)+\Delta f(n+1/2)$, where $\Delta f(n+1/2)$ is the difference between $f(n+1)$ and $f(n)$, and $\Delta f(x)$ is an odd function. Therefore, we have:

\begin{equation}
\begin{split}
& \sum_{n=-N_0}^{N_0-1}f(n)f(n+1)=\sum_{n=-N_0}^{N_0}f(n)  f(n+1)-f(N_0) f(N_0+1)\\
& = \sum_{n=-N_0}^{N_0}f(n)  \left[f(n)+\Delta f(n+1/2)\right]-f(N_0) f(N_0+1)\\
&= \sum_{n=-N_0}^{N_0}f^2(n) + \sum_{n=-N_0}^{N_0}f(n)\Delta f(n+1/2) - f(N_0)f(N_0+1)\\
& \approx \sum_{n=-N_0}^{N_0}f^2(n) - f(N_0)f(N_0+1),
\end{split}
\label{eqn:envelope}
\end{equation}{}

\noindent where we have used $f(n) \Delta f(n+1/2)$ approximated to an odd function when the spectrum is broad, and thus $|f(n+1)-f(n)| \ll f(n)$, and $\Delta f(n+1/2) \approx \Delta f(n)$. It is clear that when $N$ and $N_0$ are very large, the sum is dominated by the total optical power, $\sum_{n=-N_0}^{N_0}f^2(n)$, and is almost irrelevant to the function of the envelope. The mmWave power can be expressed as:
\begin{equation}
P_{f_r} = \frac{I_{DC}^{2}R_{L}}{2} \left[2-\frac{2f(N_0)f(N_0+1)}{\sum_{-N_0}^{N_0}f^2(n)}\right]^2.
\end{equation}
When $N_0 \to \infty$, $f(N_0)f(N_0+1) \ll \sum_{-N_0}^{N_0}f^2(n)$, and the power gain relative to the heterodyne detection approaches 6 dB regardless of the spectral envelope $f(n)$. It shall be noted that this result only applies to the case where the spectral envelope is symmetric and smooth, otherwise the approximation used in equation (9) will fail.

\end{footnotesize}

\medskip

{\noindent \bf Data availability.} The data that support the plots within this paper and other findings of this study are available from the corresponding author upon reasonable request.

\medskip

\noindent\textbf{Acknowledgement}

\noindent The authors thank Ligentec and VLC Photonics for resonator fabrication, Q.F. Yang at Caltech for helpful comments during the preparation of this manuscript, and gratefully acknowledge the support from the National Science Foundation and Defense Advanced Research Projects Agency (DARPA) under HR0011-15-C-0055 (DODOS). X.Y. is also supported by Virginia Space Grant Consortium.

\vspace{6pt}
\noindent \textbf{Competing interests} 

\noindent The authors declare no competing interests.

\vspace{6pt}
\noindent \textbf{Author Contributions} 

\noindent X.Y. and A.B conceived the concept. B.W. and J.S.M. performed the experiment, with assistance from K.S., M.J. and Z.Y. J.S.M., M.W. and S.E. fabricated the modified uni-traveling carrier photodiode. B.W., J.S.M., A.B. and X.Y. analyzed data. All authors contributed to the writing of the manuscript.

\vspace*{17mm}

\bibliography{references}


\end{document}